# Runup of nonlinear asymmetric waves on a plane beach

**Irina Didenkulova and Efim Pelinovsky**

Institute of Applied Physics, Nizhny Novgorod, Russia

**Tarmo Soomere**

Institute of Cybernetics, Tallinn, Estonia

**Narcisse Zahibo**

University of Antilles and Guyane, Guadeloupe, France

The problem of the long wave runup on a beach is discussed in the framework of the rigorous solutions of the nonlinear shallow-water theory. The key and novel moment here is the analysis of the runup of a certain class of asymmetric waves, the face slope steepness of which exceeds the back slope steepness. Shown is that the runup height increases when the relative face slope steepness increases whereas the rundown weakly depends on the steepness. The results partially explain why the tsunami waves with the steep front (as it was for the 2004 tsunami in the Indian Ocean) penetrate deeper into inland compared with symmetric waves of the same height and length.

## 1. Introduction

The reliable estimate of the extension of the flooding zone is a key problem of the tsunami prevention and mitigation. Since the characteristic length of a tsunami wave in the coastal zone is several kilometres, the nonlinear shallow water theory is an appropriate theoretical model to describe the process of the tsunami runup on the beach. Carrier and Greenspan (1958) first obtained rigorous mathematical results for the runup problem. They solved the nonlinear shallow-water equations for the case of 1+1 dimensions and a plane beach of constant slope. They applied the hodograph (Legendre) transformation to reduce the initial nonlinear hyperbolic equations in the spatial domain with an unknown boundary (*resp*. moving shoreline) to the linear wave equation on a fixed semi-axis.

After this pioneering study, also Spielfogel (1976), Pelinovsky and Mazova (1992), Tinti and Tonini (2005) have found some particular exact explicit analytical solutions to this problem for specific beach profiles and/or types of incoming waves. The main difficulty in this problem is the implicit form of the hodograph transformation. For that reason the detailed analysis of runup characteristics usually requires numerical methods. Various shapes of the periodic incident wave trains such as the sine wave (Kaistrenko et al, 1991), cnoidal wave (Synolakis, 1991), and bi-harmonic wave (Didenkulova and Kharif, 2005) have been analyzed in literature. The relevant analysis has been also performed for a variety of solitary waves and single pulses such as soliton (Pedersen and Gjevik, 1983; Synolakis, 1987), sine pulse (Mazova et al, 1991), Lorentz pulse (Pelinovsky and Mazova, 1992), gaussian pulse (Carrier et al, 2003), and *N*-waves (Tadepalli and Synolakis, 1994). In particular, antisymmetric disturbances such as *N*-waves are considered now as the realistic initial conditions of the earthquake-forced tsunamis (Tadepalli and Synolakis, 1996; Tinti and Tonini, 2005).

It is well known that nonlinear long wave evolution in shallow water even of constant depth results in the deformation of the wave profile and, finally, to the wave breaking (see, for instance, Stoker, 1957: Whitham, 1974; Voltsinger et al, 1989; Tan, 1992; Kapinski, 2006). A tsunami wave is not an exception. It usually propagates over a long distance and, even if originally perfectly symmetric and linear entity, its shape is eventually modified due to nonlinearity. The increase of the steepness of the tsunami wave front is predicted theoretically in (Murty, 1977; Pelinovsky, 1982) and is reproduced in the numerical simulation of the tsunami over long distances (Zahibo et al, 2006). There are a lot of observations of the wave breaking and its transformation into the undular bore made during the huge tsunami in the Indian Ocean on $26^{th}$ December 2004. Analogous processes are commonly observed when tsunami waves enter an estuary or a river mouth (Pelinovsky, 1982; Tsuji et al, 1991), or penetrate into straits or channels (Pelinovsky and Troshina, 1994; Wu and Tian, 2000; Caputo and Stepanyants, 2003).

The main goal of this paper is to demonstrate the significant increase of the runup height in the particular case when the incoming wave has a steep front compared with the symmetric waves with the same parameters. This effect will be studied using the exact solutions of the nonlinear shallow-water equations. We also present a simple algorithm of calculating the conditions of the wave breaking (so-called gradient catastrophe) without using the Jacobian of the hodograph transformation. The paper is organized as follows. The method of solution of the runup problem in the framework of the nonlinear shallow-water theory based on the hodograph transformation is described in section 2. Matching of the runup zone with the shelf of constant depth, and the nonlinear transformation of the shallow-water wave above even bottom is considered in section 3. The runup of the nonlinear asymmetric waves is studied in section 4. Main results are summarized in section 5.

## 2. Mathematical Model and Hodograph Transformation

The classical nonlinear shallow water equations for 2D water waves in the ideal fluid with linearly sloping bottom (Fig. 1) are:

$$\frac{\partial \eta}{\partial t} + \frac{\partial}{\partial x}\left[(-\alpha x + \eta)u\right] = 0, \qquad (1)$$

$$\frac{\partial u}{\partial t} + u\frac{\partial u}{\partial x} + g\frac{\partial \eta}{\partial x} = 0, \qquad (2)$$

where $\eta(x,t)$ is the surface displacement, $u$ is the depth-averaged velocity, $g$ is the gravity acceleration, and $\alpha$ is the bottom slope.

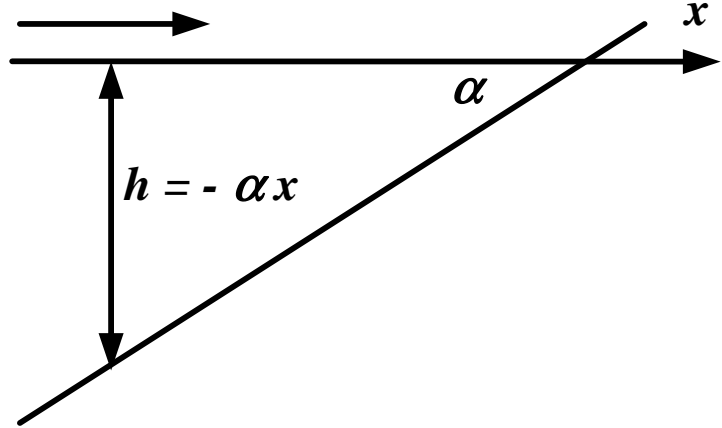

**Fig. 1.** Definition sketch for the wave runup problem

It is convenient to rewrite Eqs. (1, 2) through their Riemann invariants and to apply the hodograph transformation to the resulting equations (Carrier and Greenspan, 1958). Doing so leads to the linear wave equation with respect to the wave function $\Phi$

$$\frac{\partial^2 \Phi}{\partial \lambda^2} - \frac{\partial^2 \Phi}{\partial \sigma^2} - \frac{1}{\sigma}\frac{\partial \Phi}{\partial \sigma} = 0, \qquad (3)$$

where the new (generalised) coordinates $\lambda$ and $\sigma$ have been introduced and all variables can be expressed through the wave function $\Phi(\sigma,\lambda)$ as follows:

$$\eta = \frac{1}{2g}\left(\frac{\partial \Phi}{\partial \lambda} - u^2\right), \qquad (4)$$

$$u = \frac{1}{\sigma}\frac{\partial \Phi}{\partial \sigma}, \qquad (5)$$

$$t = \frac{1}{\alpha g}\left(\lambda - \frac{1}{\sigma}\frac{\partial \Phi}{\partial \sigma}\right), \qquad (6)$$

$$x = \frac{1}{2\alpha g}\left(\frac{\partial \Phi}{\partial \lambda} - u^2 - \frac{\sigma^2}{2}\right), \tag{7}$$

Since

$$\sigma = 2\sqrt{g(-\alpha x + \eta)}, \tag{8}$$

and the point $\sigma = 0$ corresponds to the moving shoreline, it is sufficient to solve wave equation (3) on the semi-axis ($0 \leq \sigma < \infty$) with some initial or boundary conditions offshore. The dynamics of the moving shoreline is an extremely important feature of the flooding zone when tsunami waves approach to the coast. Its analysis for a class of asymmetric waves is the main goal of our study.

Similarly, linear equations of the shallow water theory

$$\frac{\partial \eta}{\partial t} + \frac{\partial}{\partial x}\left[(-\alpha x)u\right] = 0, \quad \frac{\partial u}{\partial t} + g\frac{\partial \eta}{\partial x} = 0 \tag{9}$$

can be also reduced to the linear wave equation (3), solution of which we call the "linear" (wave) function $\Phi_l(\sigma_l, \lambda_l)$ below. The difference compared with the above analysis consists in the use of the linear version of the hodograph transformation

$$\eta_l = \frac{1}{2g}\left(\frac{\partial \Phi_l}{\partial \lambda_l}\right), \quad u_l = \frac{1}{\sigma_l}\frac{\partial \Phi_l}{\partial \sigma_l}, \quad t_l = \frac{\lambda_l}{\alpha g}, \quad x_l = -\frac{\sigma_l^2}{4\alpha g}. \tag{10}$$

As above, Eq. (3) should be solved on a semi-axis. The point $\sigma_l = 0$ now corresponds to the unperturbed shoreline $x = 0$.

A long wave of small amplitude propagating in a deep open sea area is usually almost perfectly linear and can be described by linear theory with very high accuracy. For such an incident wave the boundary conditions for the "nonlinear" and "linear" wave equations coincide provided they are defined in a deep enough area. Consequently, the solutions of the nonlinear and linear problems also coincide and $\Phi(\sigma, \lambda) = \Phi_l(\sigma_l, \lambda_l)$. Moreover, if the "linear" solution $\Phi_l(\sigma_l, \lambda_l)$ is known, the solution of the nonlinear problem (1, 2) can be directly found from expressions (4)-(7). In fact, it is difficult to do analytically but very easy numerically. In particular, description of properties of the moving shoreline $\sigma(x,t)$ is straightforward. From (5), (6) and (10) it follows that

$$u(\lambda) = \lambda - \lambda_l,  \qquad (11)$$

or, in an equivalent form,

$$u(\lambda) = u(\lambda_l + u_l),  \qquad (12)$$

which demonstrates that the speed of the shoreline displacement can be found through the Riemann transformation of time. As the functional forms of the "linear" and "nonlinear" solutions are identical, we may re-write (12) finally:

$$u(t) = U\left(t + \frac{u}{\alpha g}\right),  \qquad (13)$$

where $U(t)$ stands for the "linear" speed of the shoreline.

Thus, if the approaching wave is linear, a rigorous "two-step" method can be used to calculate the runup characteristics. Firstly, the wave properties on the unperturbed shoreline $x=0$ such as the vertical displacement $Y(t)$ or the velocity of wave propagation

$$U(t) = \frac{1}{\alpha}\frac{dY(t)}{dt},  \qquad (14)$$

are determined within the linear problem. Its solution can be found using traditional methods of the mathematical physics. Secondly, the properties of the solution to the nonlinear problem are found from expressions derived above. For example, the real "nonlinear" speed of the moving shoreline is found from (13), and finally, the vertical displacement of the water level and position of the shoreline at some time instant horizontal distance of the flooding (*resp*. the width of the flooded area) as

$$y(t) = \alpha x(t), \quad x(t) = \int u(t)dt.  \qquad (15)$$

Using (4), Eq. (15) can be re-written as

$$y(t) = \eta(t, \sigma = 0) = Y\left(t + \frac{u}{\alpha g}\right) - \frac{u^2}{2g}. \qquad (16)$$

The important conclusion from expressions (13) and (16) is that the maxima of vertical displacements (equivalently, the runup or rundown height) and the velocity of the shoreline displacement in the linear and nonlinear theories coincide. Consequently, the linear theory adequately describes the runup height which is an extremely important characteristic of tsunami action on the shore. In fact, this conclusion was reached in many papers cited above for various shapes of the incident wave. The rigorous proof demonstrated here follows the work by Pelinovsky and Mazova (1992).

There are no rigorous results in the nonlinear theory in the case of more complicated bottom profiles that cannot be approximated by the idealized beach of constant slope. Yet the linear theory can be used in some cases when the nearshore has such a slope alone. If the wave is nonlinear only at the runup stage, the linear theory frequently can be used to describe wave transformation in the ocean of variable depth and the resulting wave can be matched with the nonlinear solutions. This approach is quite popular, see (Kanoglu, 2004) and references therein. Nonlinear effects in the transition zone (between the offshore and the runup zone) can be accounted for as the correction term to the boundary conditions far from the shoreline (Li and Raichlen, 2001).

Another important outcome from proposed approach is the simple definition of the conditions of the first breaking of the waves on a beach. It is evident that long small-amplitude waves will not break at all and result in a slow rise of the water level resembling surge-like flooding. With increase of the wave amplitude, the breaking appears seawards from the runup maximum and, depending on the wave amplitude and the bottom slope, may occur relatively far offshore. The above approach allows determining the position of the first breaking from the expressions for the dynamics of the shoreline. The temporal derivative of the velocity of the moving shoreline, found from (13),

$$\frac{du}{dt} = \frac{dU/dt}{1 - \dfrac{dU/dt}{\alpha g}}, \qquad (17)$$

tends to the infinity (equivalently, wave breaking occurs, Pelinovsky and Mazova 1992) when the denominator a the right-hand side of Eq. (17) approaches to zero. The condition of the first wave breaking therefore is

$$Br = \frac{\max(dU/dt)}{\alpha g} = \frac{\max(d^2Y/dt^2)}{\alpha^2 g} = 1. \tag{18}$$

This condition has a simple physical interpretation: the wave breaks if the maximal acceleration of the shoreline $Y''\alpha^{-1}$ along the sloping beach exceeds the along-beach gravity component ($\alpha g$). This interpretation is figurative, because formally $Y''$ only presents the vertical acceleration of the shoreline in the linear theory and the "nonlinear" acceleration $du/dt$ actually tends to infinity at the breaking moment.

The above-cited literature contains various examples of studies of long wave runup on the plane beach using the hodograph transformation. The effectiveness of this two-step approach can be demonstrated by considering the runup of a sine wave with frequency ω. The well-known bounded solution of the linear wave equation (3) is expressed in the Bessel functions

$$\eta(x,t) = R J_0\left(\sqrt{\frac{4\omega^2 |x|}{g\alpha}}\right)\cos(\omega t), \tag{19}$$

where $R$ is the maximal wave amplitude at the unperturbed shoreline $x = 0$. As mentioned above, it is also the maximal runup height in the nonlinear theory. Far from the shoreline the wave field can be presented asymptotically as the superposition of two sine waves of equal amplitude propagating in the opposite directions

$$\eta(x,t) = A(x)\left\{\sin\left[\omega(t-\tau)+\frac{\pi}{4}\right]+\sin\left[\omega(t+\tau)-\frac{\pi}{4}\right]\right\}, \tag{20}$$

where the instantaneous wave amplitude $A(x)$ is

$$A(x) = R\left(\frac{\alpha g}{\pi^2 \omega^2 |x|}\right)^{1/4}, \tag{21}$$

and the propagation time of this wave over some distance in a fluid of variable depth is

$$\tau(x) = \int \frac{dx}{\sqrt{gh(x)}}. \tag{22}$$

The maximum change of the amplitude of the approaching wave with the wavelength $\lambda$ (determines from dispersion relation, $\omega = ck$) and the initial amplitude $A_0$ at the fixed point $|x| = L$ is characterized by the amplification factor (equivalent to the shoaling coefficient in the linear surface wave theory), which can be found from (21):

$$\frac{R}{A_0} = \left(\frac{\pi^2 \omega^2 L}{g\alpha}\right)^{1/4} = 2\pi\sqrt{\frac{2L}{\lambda}}. \tag{23}$$

We emphasize that the amplification factor in (23) calculated in the framework of the linear theory is the same in the nonlinear theory. This feature allows to determine the extreme runup characteristics in both cases if the initial wave amplitude and length are known. Using (23), the extreme values for the velocity of the moving shoreline and the breaking criterion can be calculated as follows:

$$U_{ext} = \frac{\omega R}{\alpha}, \tag{24}$$

$$Br_{\sin} = \frac{\omega^2 R}{g\alpha^2} = 1. \tag{25}$$

As a result, it is simple to predict the minimal value (threshold) of the runup height when the wave breaking appears from any given wave frequency (or period) and the bottom slope.

### 3. Nonlinear Wave Deformation

An adequate theory of runup should take into account the potential asymmetry of the incoming waves. Such waves appear, for example, when the shelf offshore from the runup zone has a flat bottom (Fig. 2). In the water of constant depth, the exact one-wave solution of the nonlinear shallow water equations can be described by the following partial differential equation, which can be easily derived from (1)-(2):

$$\frac{\partial \eta}{\partial t} + V\frac{\partial \eta}{\partial x} = 0,$$

(26)

$$V = \sqrt{gh} + \frac{3u}{2} = 3\sqrt{g(h+\eta)} - 2\sqrt{gh}, \qquad u = 2\left(\sqrt{g(h+\eta)} - \sqrt{gh}\right).$$

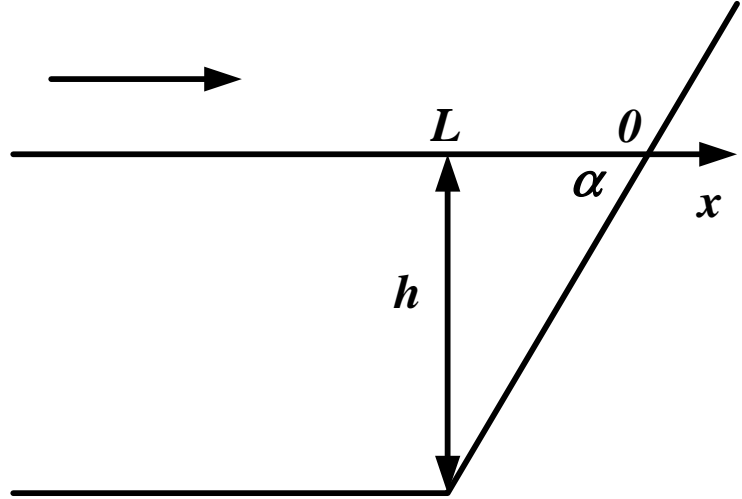

**Fig. 2.** Definition sketch of the coastal geometry

The solution of Eqs. (26) satisfying the initial condition $\eta(x,t=0) = \eta_0(x)$ is the Riemann wave

$$\eta(x,t) = \eta_0(x - Vt).$$

(27)

Its shape varies with distance and its steepness increases due to the difference in speed of the crest and trough. The instantaneous slope of the water surface at any point of the incoming wave is

$$\frac{\partial \eta}{\partial x} = \frac{\eta_0'}{1 + tV_0'},$$

(28)

where the prime means $d/d\tilde{x}$, where $\tilde{x} = x - Vt$, and $V_0(x)$ is determined through the initial wave shape $\eta_0(x)$ with use of (27). On the face of the incident wave $\partial \eta/\partial x < 0$, $\partial V_0/\partial x < 0$ and the denominator at the right-hand side of (28) decreases with time; thus the wave steepness increases and becomes infinite at

$$t = T = \frac{1}{\max(-V_0')}. \qquad (29)$$

As an example, we analyse the nonlinear deformation of the initial sine wave with an amplitude *a* and a wave number *k* propagating in water of constant depth. The temporal evolution of the wave shape is demonstrated in Fig. 3 for several initial dimensionless amplitudes *a/h* (Stoker, 1957; Whitham, 1974; Pelinovsky, 1982; Kapinski, 2006; Didenkulova et al, 2006). The breaking time and the breaking distance *X* are

$$X = \sqrt{gh}T = \frac{1}{3k}\sqrt{\frac{2}{1-\sqrt{1-(a/h)^2}}}. \qquad (30)$$

The breaking distance decreases when the wave amplitude increases. Large-amplitude waves break almost after their generation, but waves with small amplitudes may pass a long distance before breaking.

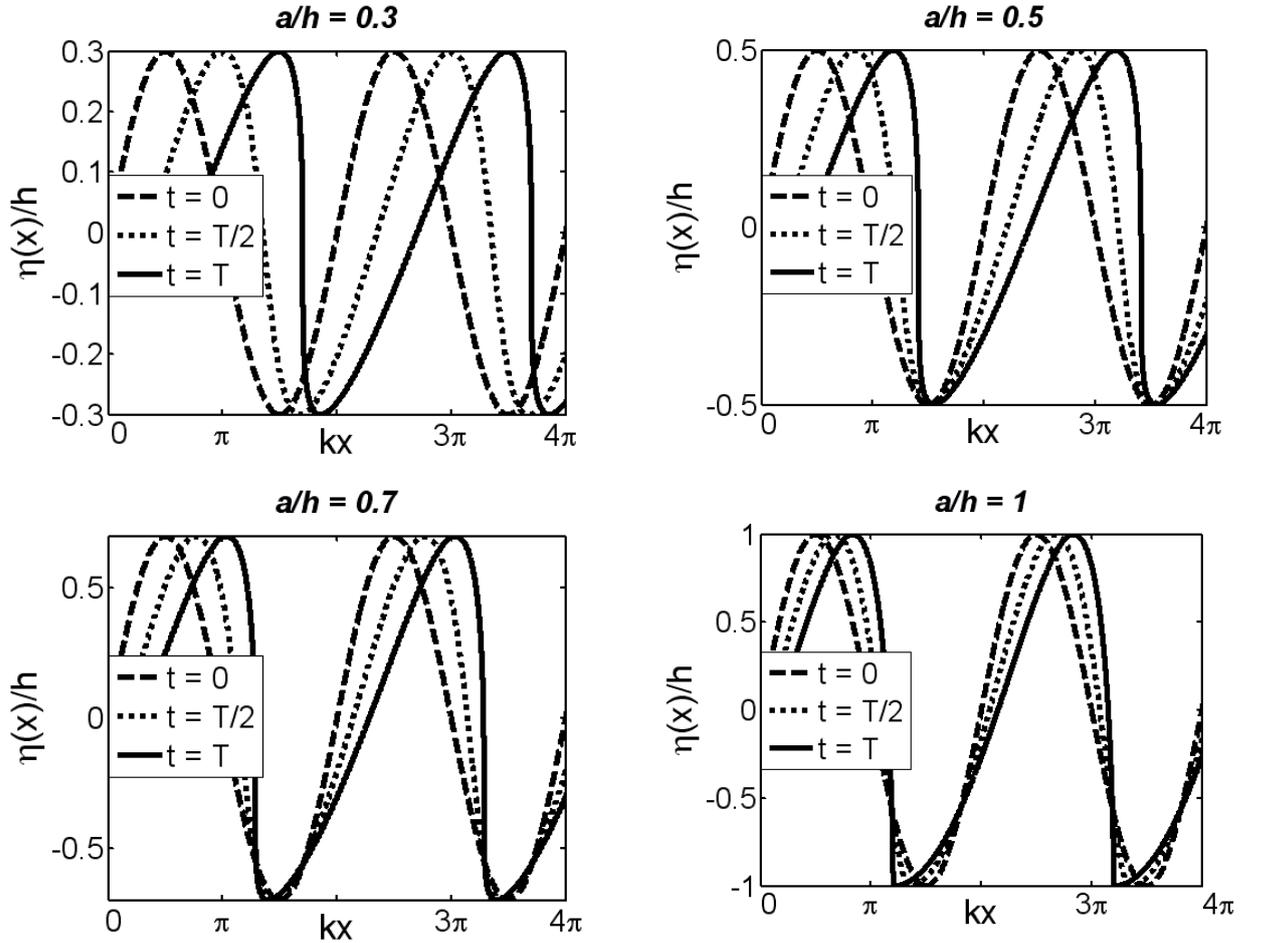

**Fig. 3.** Deformation of an initial sine wave (dashed line) in shallow water

The following simplified formula for the maximum steepness of the face-slope (Fig. 4) can be derived from (29):

$$s = \max(\partial \eta / \partial x) = \frac{s_0}{1 - t/T}, \quad (31)$$

where $s_0 = ak$ is the initial wave steepness. At the breaking time, the steepness is infinite.

For certain applications it is important to know the spectrum of the shallow-water wave. The spectral presentation of the Riemann wave in terms of sine harmonics can be presented explicitly (Pelinovsky, 1982; Didenkulova et al, 2006):

$$\eta(t, x) = \sum_{n=1}^{\infty} A_n(t) \sin\left(nk[x - \sqrt{ght}]\right), \quad A_n(t) = 2a \frac{T}{nt} J_n\left(\frac{nt}{T}\right), \quad (32)$$

where $J_n$ are the Bessel functions. The amplitudes of the higher harmonics increase with time whereas the amplitude of the basic harmonic corresponding to $n = 1$ decreases (Fig. 5).

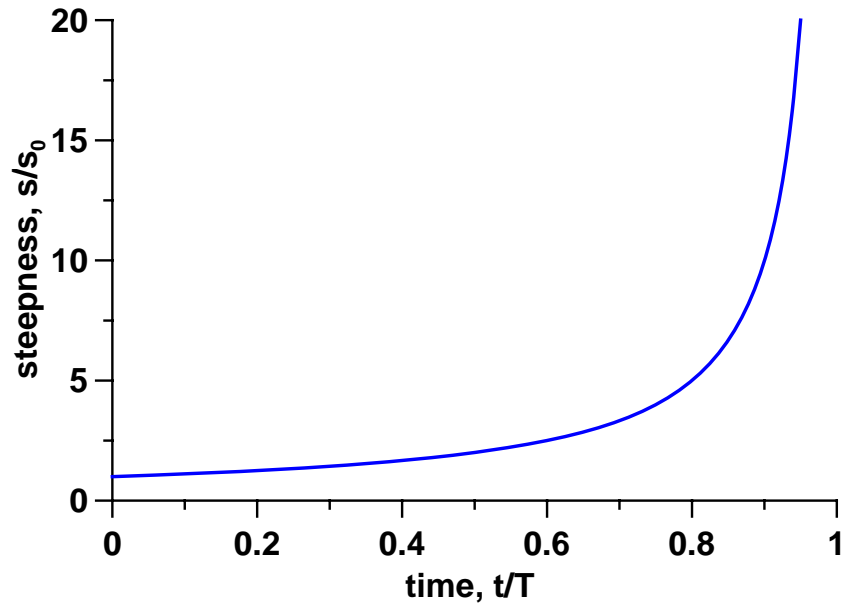

**Fig. 4.** Temporal evolution of the wave steepness

The realistic tsunami wave evolution both in the open sea and in the coastal zone is extremely complicated due to effects of refraction, diffraction and resonance. The propagation time, used to characterize the wave properties in the simple example of nonlinear wave deformation considered above, technically may be used also in the general case. However, in the general case, the wave steepness is a more convenient measure of the wave shape than the propagation time. Using (31), the spectral amplitudes (32) can be expressed as

$$A_n(s) = \frac{2a}{n(1-s_0/s)} J_n\left(n\left[1-\frac{s_0}{s}\right]\right). \tag{33}$$

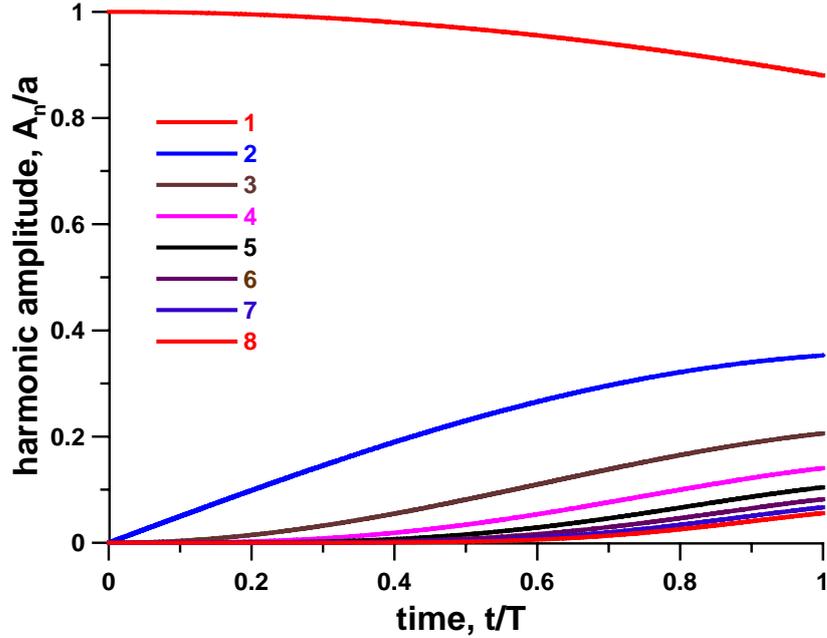

**Fig. 5.** Temporal evolution of the amplitude of the spectral harmonics

Since the spectral amplitudes with $n>1$ increase with the steepness increasing, we may determine the relation between "local" or "current" wave characteristics without considering the wave history.

## 4. Runup of Nonlinear Deformed Wave on a Plane Beach

The nonlinear long wave propagation in a large ocean of constant depth is thus always accompanied by a certain deformation of the wave shape. Such wave coming to the beach of constant slope (Fig. 2) has a front, much steeper, than its back. The runup of such asymmetric waves on the plane beach can be studied with the use of the model described in section 2.

The first step the two-step approach consists in solving the linear problem. For doing this we may use the linear superposition of elementary solutions (19) and match it with the Riemann wave (32) far from the shoreline. If the far-field wave in (20) is approximated by a (finite or infinite) superposition of harmonics, the use of expressions (21)-(23) leads to the following expressions for the incident wave on a distance $L$ from the shoreline where the beach is matched with a shelf of constant depth and for the "linear" oscillations of the water level at the unperturbed shoreline:

$$\eta(t, x = L) = \sum_{n=1}^{\infty} A_n(s) \sin(n\omega t]), \quad A_n(s) = \frac{2a}{n(1-s_0/s)} J_n\left(n\left[1-\frac{s_0}{s}\right]\right), \quad (34)$$

$$Y(t) = \eta(t, x = 0) = \left(\frac{\pi \omega L}{gh}\right)^{1/2} \sum_{n=1}^{\infty} \sqrt{n} A_n(s) \sin\left[n\omega(t-\tau) + \frac{\pi}{4}\right], \quad (35)$$

where $A_n$ is the amplitude of the $n$-th harmonic, $\omega = k\sqrt{gh}$ is the frequency of the initial sine wave, and $\tau$ is the time of wave propagation from $x = L$ to $x = 0$. It is convenient to normalize water level oscillations at the shoreline against the runup height for the sine wave (23). The expression for the normalized water level oscillations at the shoreline is

$$Y^*(t^*) = \sum_{n=1}^{\infty} \sqrt{n} \frac{A_n}{a} \sin\left(nt^* + \frac{\pi}{4}\right), \quad (36)$$

where $t^* = \omega(t-\tau)$. For convenience the asterisk will be omitted in what follows. The above has shown that the extreme values of $Y(t)$ correspond to the maximal runup height and rundown dropdown in the nonlinear theory, the dimensionless values of which (Fig. 6) are the functions of the wave steepness only. The rundown amplitude depends from the wave steepness weakly (no more then 30%), and we may use expression (23) to evaluate its approximate value. In the contrary, the runup height fast increases when the wave steepness increases. It tends to infinity for a shock wave that theoretically can be described by the model in question where the waves are assumed to be non-breaking. In realistic conditions, of course, the wave breaking will bound the runup height. The maximum runup height can be roughly approximated by (in dimension variables)

$$R_{max} = 2\pi a \sqrt{\frac{2Ls}{\lambda s_0}}. \quad (37)$$

Expression (37) shows that the wave steepness is the most significant parameter of the runup process. Further, expression (37) confirms that from all the waves of a fixed height and length from the class of waves in question, the wave with steepest front penetrates inland to the largest distance, and that all asymmetrical waves with the front steeper than the back create a larger flooding then a wave with a symmetrical shape. Many examples of extremely large penetration of tsunami waves to inland (including observations during the 2004 Indonesian tsunami) can be interpreted as the confirmation of the important role of the wave steepness.

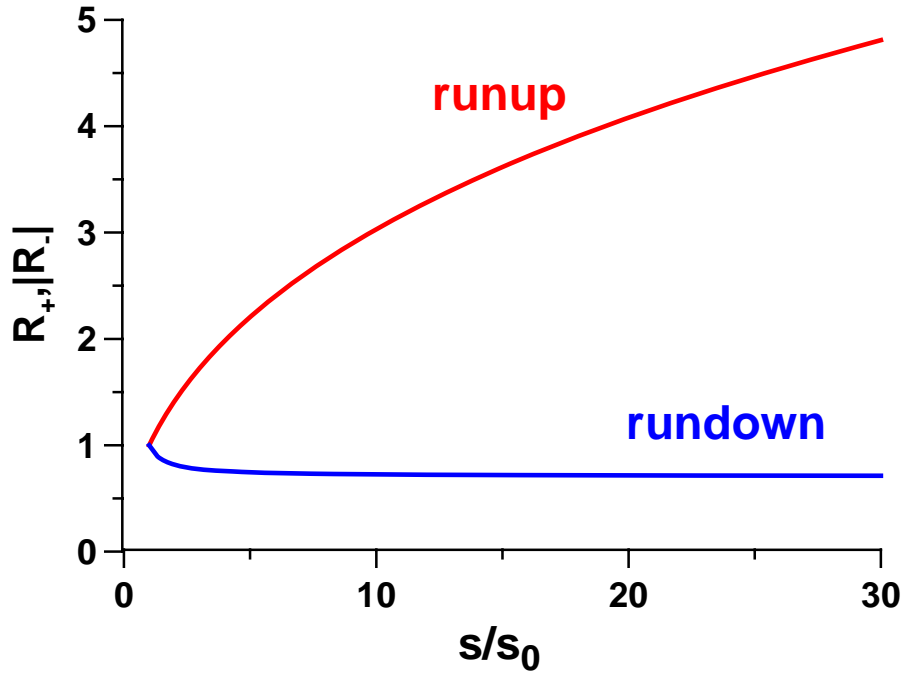

**Fig. 6.** Runup ($R_+$) and rundown ($R_-$) amplitudes versus the wave steepness

Similar analysis can be performed for extreme properties of the shoreline velocity. The dimensionless expression for the "linear" velocity (normalized against the runup velocity for sine wave) is

$$U(t) = \sum_{n=1}^{\infty} \sqrt{n^3} \frac{A_n}{a} \sin\left(nt + \frac{3\pi}{4}\right). \tag{38}$$

The extreme values of this function correspond to the "nonlinear" maximal values of runup and rundown velocity of the moving shoreline. The runup velocities exceed the rundown velocity (Fig. 7).

The runup velocity can be approximated (in dimensional variables) by

$$U_{max} = (2\pi)^2 a \sqrt{\frac{g}{h}} \left(\frac{Ls}{\lambda s_0}\right)^{3/2}. \tag{39}$$

The linear theory of long wave runup allows also to estimate the parameters of the wave breaking that occurs when

$$Br = Br_{\sin} \max\left(\frac{d^2Y}{dt^2}\right) = Br_{\sin} \max\left(\frac{dU}{dt}\right) = 1, \qquad (40)$$

where $Br_{\sin}$ is the breaking criterion for sine wave (25), and all derivatives in (40) are calculated using dimensionless expressions (36) and (38). The curve defined by Eq. (40) on plane ($Br_{\sin}$, $s/s_0$) separates the surging and plunging scenarios of wave runup (Fig. 8). The breaking scenario is, as expected, more typical for asymmetric waves with a relatively steep front.

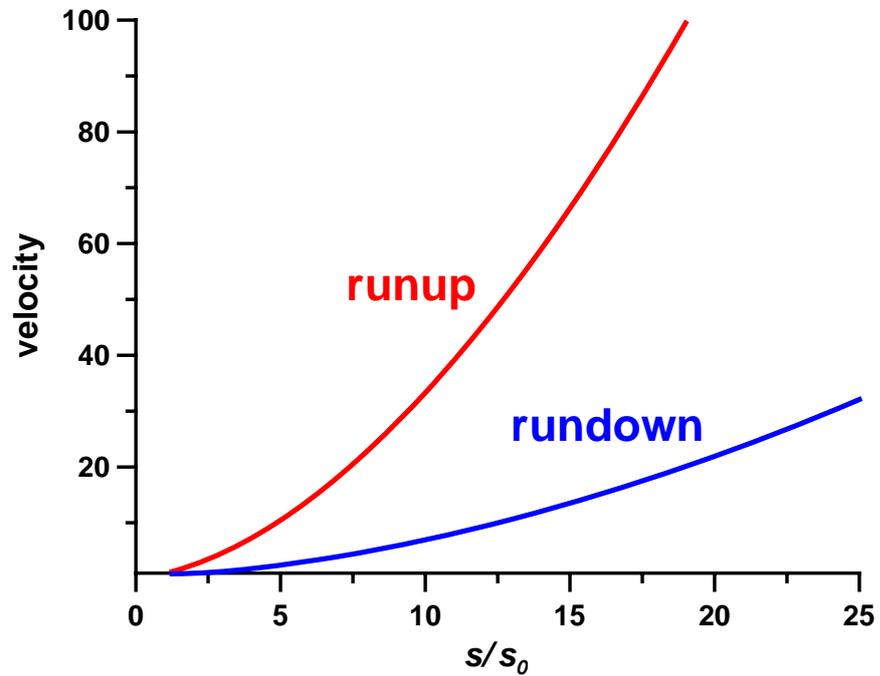

**Fig. 7.** Extreme runup and rundown velocities versus the wave steepness

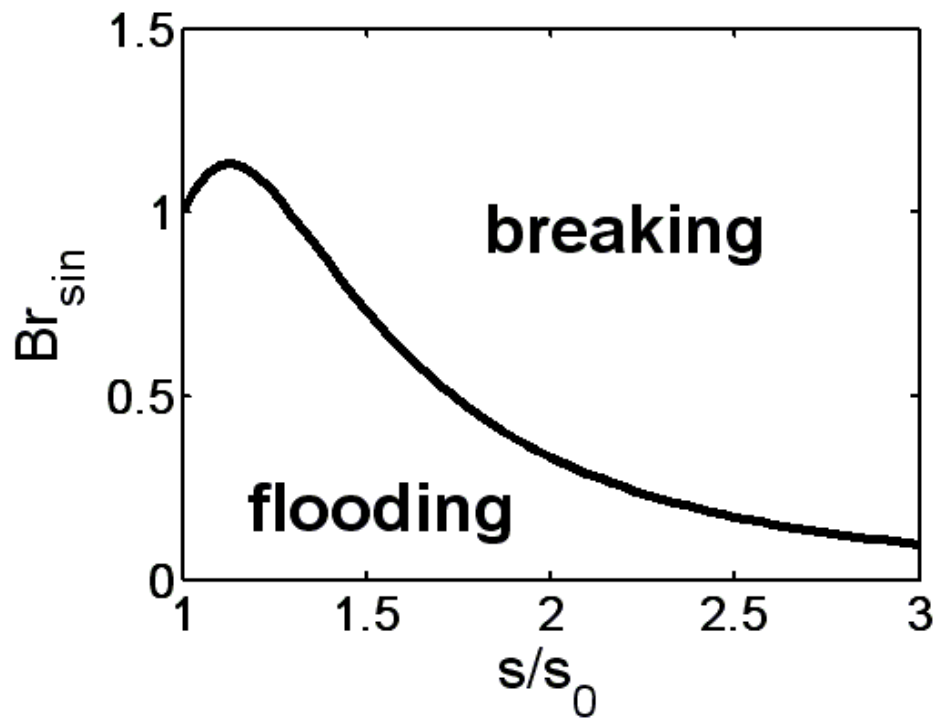

**Fig. 8.** Various scenarios of the wave runup on a beach

The second step of solving the nonlinear runup problem consists in transformation of the "linear" expressions (36) and (38) for the water level displacement and velocity into "nonlinear" expressions for the moving shoreline with use of (13) and (15). Fig. 9 displays the "nonlinear" and "linear" time history of the water level and velocity of the moving shoreline (in dimensionless form) for a symmetrical sine incident wave ($s=s_0$).

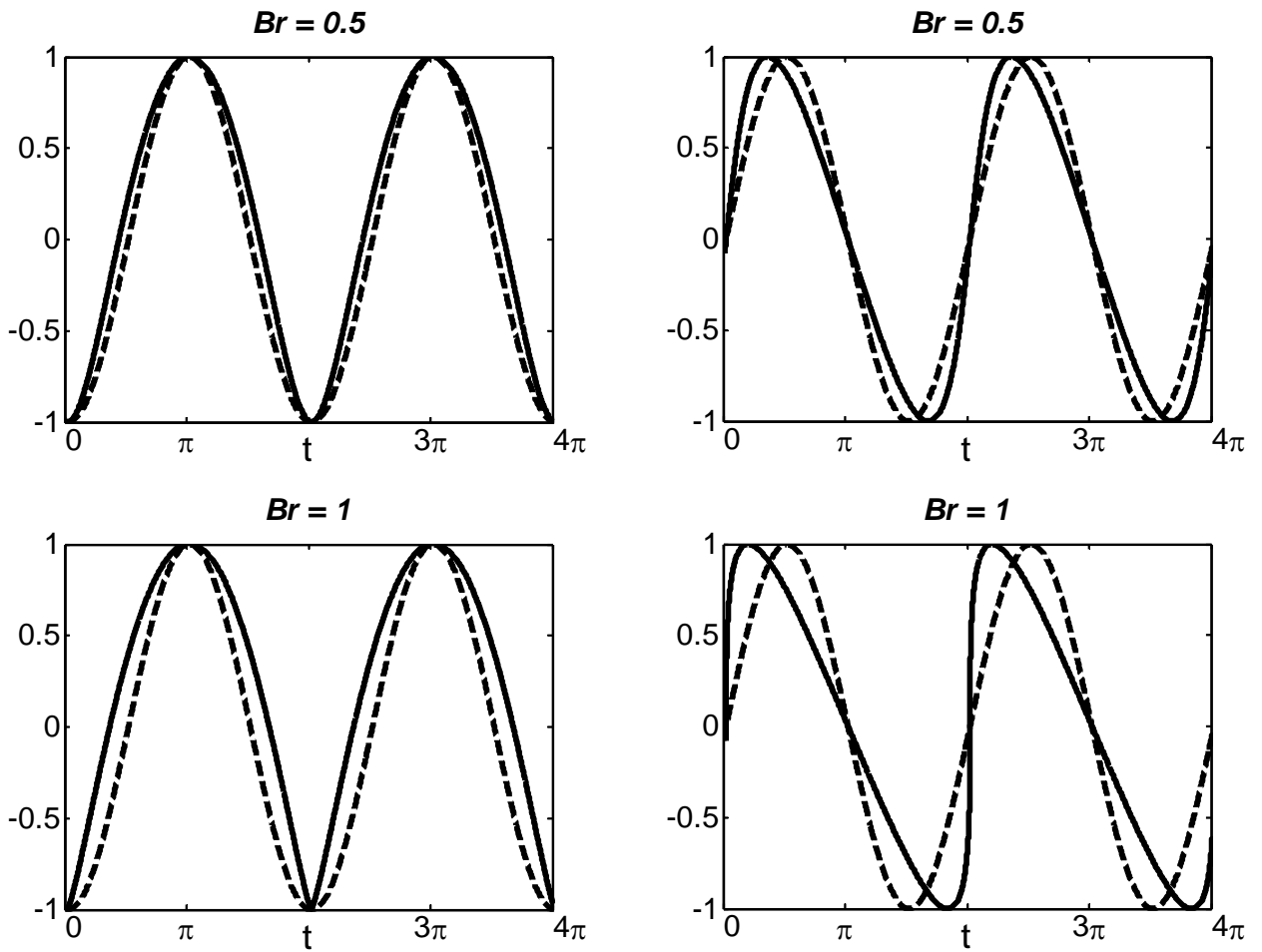

**Fig. 9.** Water level (left) and velocity (right) of the moving shoreline for various values of the breaking parameter and for a sinusoidal incident wave ($s = s_0$). Solid line corresponds to the "nonlinear" and dashed line to the "linear" solution.

If the wave is small ($Br << 1$), the position of the shoreline varies almost sinusoidally. An increase of the wave amplitude (equivalently, an increase of $Br$) in the nonlinear case is accompanied by forming of a region where the velocity changes very fast (equivalently, its graphical representation has a very steep front). The water surface displacement record tends to behave as a parabolic function; meanwhile the corresponding "linear" characteristics of the wave are sine functions. The first breaking appears at the stage of maximal rundown. The runup of the sine wave is described in (Carrier and Greenspan, 1958; Pelinovsky and Mazova, 1992) and here reproduced for illustration.

The runup of an asymmetric wave is greatly different from the runup of the sine wave. The relevant results for the case $s = 2s_0$ are presented in Fig. 10. The temporal behaviour of both the position of the shoreline and the time record of velocity are asymmetric even when the wave amplitude is small. The runup amplitude and velocity of the shoreline displacement is higher than the rundown amplitude and the relevant velocity. The breaking point is located closer to the unperturbed shoreline than in the case of the sine wave runup.

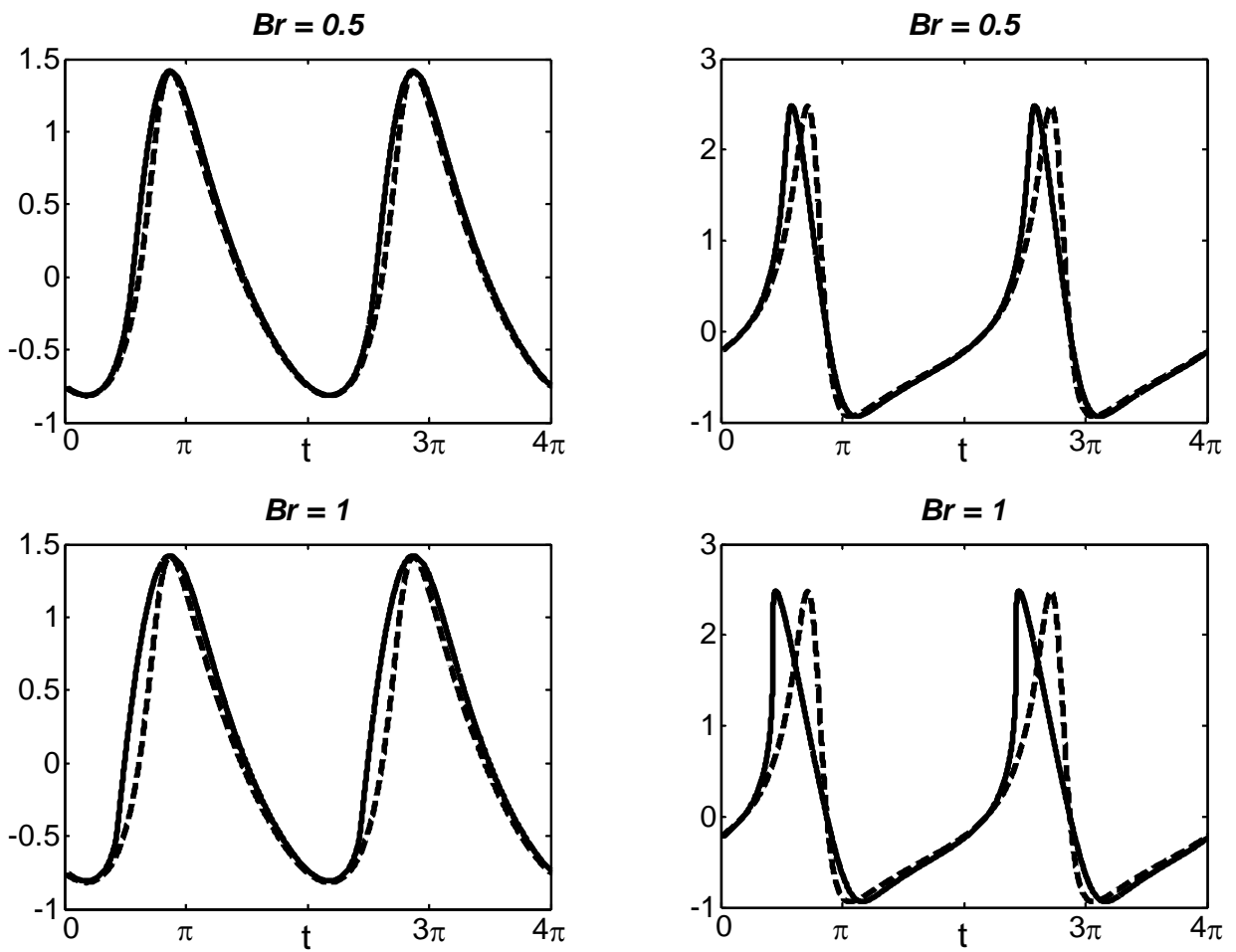

**Fig. 10.** Water level (left) and velocity (right) of the moving shoreline for various values of the breaking parameter for a moderately asymmetric incident wave ($s = 2s_0$). Solid line corresponds to the "nonlinear" and dashed line to the "linear" solution.

If the incident wave is strongly asymmetric wave ($s = 10s_0$, Fig. 11), the strong flow moves inland during a short time. The runup amplitude is higher than the rundown amplitude. Such intense flows can be distinguished on many images of the catastrophic 2004 tsunami in the Indian Ocean and eventually occurred in many sections of the affected coastline. In this case the incident wave is extremely steep and breaks rapidly, and there is almost no difference in the "linear" and "nonlinear" results.

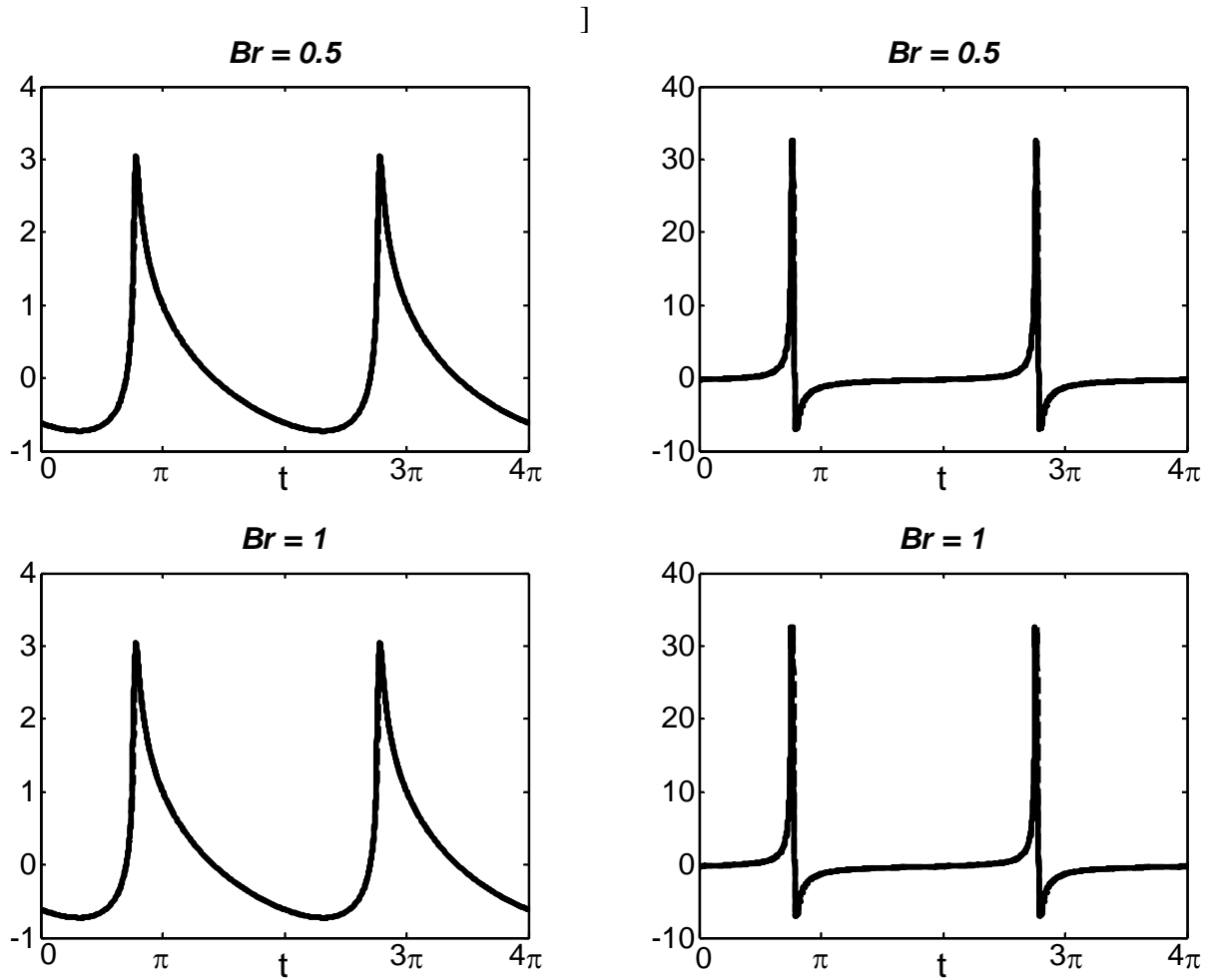

**Fig. 11.** Water level (left) and velocity (right) of the moving shoreline for various values of the breaking parameter for a strongly asymmetric incident wave ($s = 10s_0$). Solid line corresponds to the "nonlinear" and dashed line to the "linear" solution.

## 5. Conclusion

The principal result of this study is the strong influence of the wave steepness on the runup characteristics of the long waves in the framework of the analytical theory of the nonlinear shallow-water waves. Among waves of a fixed amplitude and frequency (length), the steepest wave penetrates to inland to the largest distance and with largest speed. Consequently, the least dangerous is the symmetric sine wave.

This research is supported particularly by grants from INTAS (03-51-4286) and RFBR (05-05-64265) for ID and EP; University of Antilles and Guyane for EP and NZ, Marie Curie network SEAMOCS (MRTN-CT-2005-019374) for ID, and ESF grant 5762 for TS.